\documentclass[english,english,showpacs,amsmath,prb,floatfix]{revtex4}
\usepackage[T1]{fontenc}
\usepackage[latin1]{inputenc}
\usepackage{amsmath}
\usepackage{graphicx}
\usepackage{amssymb}

\makeatletter

\providecommand{\tabularnewline}{\\}

\usepackage{float}

\usepackage{array}\usepackage{float}\usepackage{rotfloat}

\usepackage{babel}

\makeatother

\usepackage{babel}

\makeatother

\usepackage{babel}

\usepackage{babel}
\makeatother

\begin{document}

\title{Energetics and electronic structure of phenyl-disubstituted polyacetylene:
A first-principles study}

\author{Priya Sony$^{\text{1,2*}}$}

\author{Alok Shukla$^{\text{2}}$}

\author{Claudia Ambrosch-Draxl$^{\text{1}}$ }

\affiliation{$^{\text{1}}$Chair of Atomistic Modelling and Design of Materials,
University of Leoben, Franz-Josef-Straße 18, A--8700 Leoben, Austria\\
 $^{\text{2}}$Department of Physics, Indian Institute of Technology
Bombay, Powai, Mumbai--400076, Maharashtra, India}

\email{psony11@gmail.com}

\begin{abstract}
Phenyl-disubstituted polyacetylene (PDPA) is an organic semiconductor
which has been studied during the last years for its efficient photo-luminescence.
In contrast, the molecular geometry, providing the basis for the electronic
and optical properties, has been hardly investigated. In this paper,
we apply a density-functional-theory based molecular-dynamics approach
to reveal the molecular structure of PDPA in detail. We find that
oligomers of this material are limited in length, being stable only
up to eight repeat units, while the polymer is energetically unfavorable.
These facts, which are in excellent agreement with experimental findings,
are explained through a detailed analysis of the bond lengths. A consequence
of the latter is the appearance of pronounced torsion angles of the
phenyl rings with respect to the plane of the polyene backbone, ranging
from $55^{\circ}$ up to $95^{\circ}$. We point out that such large
torsion angles do not destroy the conjugation of the $\pi$ electrons
from the backbone to the side phenyl rings, as is evident from the
electronic charge density. 
\end{abstract}


\maketitle

\section{Introduction}

In the emerging field of nanotechnology conjugated polymers are playing
an important role as materials of choice for organic light emitting
diodes (OLEDs),\citet{burroughes1990,sirringhaus1998} organic field
effect transistors (OFETs),\citet{gigli2001} organic lasers,\citet{kumar1997,moses1992}
photocells,\citet{yu1994} \emph{etc}. Their application in such devices
requires the search and the investigation of new materials with interesting
optical and electronic properties. PA which is the simplest and the
most widely studied conjugated polymer, exhibits very weak photo-luminescence\citet{burroughes1990,yoshino1983,tpa}
(PL), thus ruling it out as a candidate for opto-electronic applications.
However, in the last decade, phenyl-disubstituted polyacetylene (PDPA)
derivatives, obtained by replacing hydrogen atoms of \emph{trans}-polyacetylene
(PA) by phenyl rings or their derivatives, were demonstrated as luminescent
materials with high quantum efficiency, which also exhibited stimulated
emission in thin film form.\citet{tada-fujii-book1997,gontia1999}
Therefore, one anticipates that they will be useful in creating light
emitting diodes and polymeric lasers.\citet{shukla_PRL1999}

To understand the physics behind the fluorescence of these materials,
a series of experimental and theoretical investigations have been
performed.\citet{Fujii2001,gontia1999,gustafson2001,hidayat2001,frolov1998,shukla2001,ghosh_PRB2000,shukla_PRL1999,korovyanko2003,shukla_chemphys2004,shukla_prb2004,sony-pdpa2005,An2003}
From calculations based on a Pariser-Parr-Pople (PPP) model Hamiltonian
it was argued that this phenomenon is due to reverse excited-state
ordering (compared to PA) of the one-photon allowed $1B_{u}$ and
two-photon allowed $2A_{g}$ states, caused by reduced electron correlation
effects due to the presence of phenyl rings.\citet{ghosh_PRB2000,shukla2001,shukla_PRL1999}
In PA, the $2A_{g}$ state occurs below the $1B_{u}$ level, as a
consequence of which the optically pumped $1B_{u}$ state decays rapidly
to the $2A_{g}$ state, and according to dipole transition rules,
radiative transition between $2A_{g}$ and the $1A_{g}$ states is
forbidden. Hence it was concluded that in contrast to PA, PDPA shows
strong PL as $E(2A_{g})>E(1B_{u})$, where $1B_{u}$ is strongly dipole
coupled to the ground state.

While most of the investigations concentrated on the optical properties
of PDPAs their structural properties have hardly been investigated.
A complete understanding of the electronic and hence optical properties,
is, however, impossible without knowing the respective ground state
geometries. This leads to another puzzling issue, namely the reason
behind the experimentally observed short conjugation length of these
materials, which typically consist of seven repeat units in PDPA thin
films,\citet{Fujii2001} and only five in solution.\citet{gontia_prb66_2002}
Based upon an SSH model calculations, it was predicted in an earlier
work that for an infinite chain, introduction of phenyl rings will
lead to a reduction of bond alternation along the backbone.\citet{shukla_PRL1999}

The aim of this work is to throw light on the formation of PDPAs by
investigating their energetics as a function of oligomer length. To
this extent we carry out first-principles calculations based on density
functional theory (DFT). We thereby focus on the geometry, including
a detailed study of bond lengths and alternations, as well as the
orientation of the phenyl rings with respect to the polyene backbone.
We will show that polymerization of PDPSs is hampered by single-bond
breaking, and oligomers are only stable up to eight repeat units.
Moreover, an analysis of band gaps and charge densities allows for
an insight into the lowest optical transitions.

\section{PDPA and its oligomers}

The crystal structure of PA is of herringbone type, where the polymer
chains are coupled by the weak van der Waals interactions. Although
the 3D environment can have a large impact on the optical properties
on such systems,\citet{Puschnig2002,Hummer2004} it can be considered
as quasi one-dimensional for the current investigation, where we study
the cohesive properties as a function of chain length. Hence we model
PA, and consequently PDPA, as strictly one-dimensional zigzag chains.

\begin{figure}
\includegraphics[width=6cm]{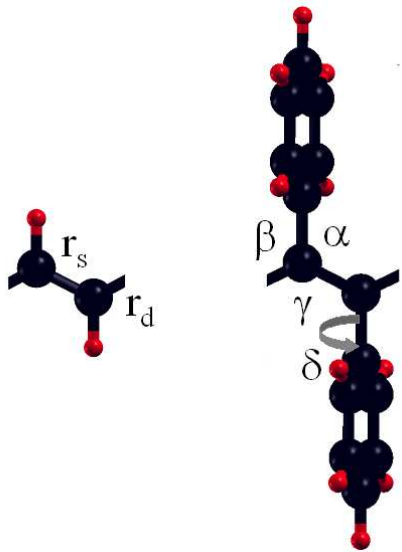}

\caption{(color online) Repeat units of polyacetylene and phenyl-disubstituted
polyacetylene. Large and small spheres denote carbon and hydrogen
atoms, respectively. $r_{s}$ represent single bonds and $r_{d}$
double bond lengths.}

\label{fig:repeat_units} 
\end{figure}

The PDPAs with the formula unit C$_{14}$H$_{10}$ are obtained by
replacing the hydrogen atoms of PA (C$_{2}$H$_{2}$) by phenyl groups,
leading to the structure presented in Fig. \ref{fig:repeat_units}.
The backbone polyene chain is running along the $x$ axis (longitudinal
direction) while the phenyl rings are oriented along the $y$ axis
(transverse direction). The single and double bonds are labeled as
$r_{s}$ and $r_{d}$, respectively, while the bond alternation is
defined as $\Delta r=r_{s}-r_{d}$. $\alpha$ and $\beta$ denote
the angles between the carbon atoms of the backbone and the phenyl
ring, $\gamma$ is the angle between the carbon atoms of the backbone,
and $\delta$ is the torsion angle of the phenyl ring with respect
to the backbone plane. The PDPA oligomers are terminated with one
hydrogen atom at either end, saturating the dangling bonds. The formula
units of PDPAs and PAs are C$_{14n}$H$_{10n+2}$ and C$_{2n}$H$_{2n+2}$,
respectively, with $n$ being a positive integer representing the
number of repeat units. We adopt the notation PDPA-$n$, and correspondingly
PA-$n$ for the oligomers of PA.

\section{Computational details}

In the present work, infinite PDPA as well as its oligomers with two
to six building blocks are investigated by means of density functional
theory employing the supercell approach. A vacuum region of around
8 \AA{}\ is adopted along the $y$ and $z$ directions for the polymers,
and also in the third dimension for the oligomers. This size is considered
sufficient to avoid interaction between the translational images.
All the geometries are optimized, where we assume the structures to
be relaxed when the forces on individual atoms are below 1 meV/\AA{}.
To this extent, the projector augmented wave (PAW) method is utilized
as implemented in the original PAW code.\citet{PAW} It makes use
of Car-Parinello molecular dynamics allowing for geometry optimization
in an accurate and efficient manner. Polynomial-type pseudopotentials
as implemented in the code are used, with a plane wave cutoff of $30$~Ry.
A Brillouin zone sampling of $16\times1\times1$ \textbf{k} points
is taken for the calculations of polymers, while the finite systems
naturally requires one \textbf{k} point only. Exchange-correlation
effects are treated by the generalized gradient approximation (GGA)
in the Perdew-Burke-Ernzerhof (PBE) flavor.\citet{PBE} We further
discuss the issue of our choice of the exchange-correlation functional
at the end of section \ref{sub-sec:structure}.

The polymerization energy $E_{poly}$ is calculated using the formula
\begin{equation}
E_{poly}=E_{polymer}-E_{monomer}\label{eq:polarization_energy}\end{equation}
 where $E_{polymer}$ is the total energy per unit cell of the infinite
polymer and $E_{monomer}$ is that of diphenyl acetylene, C$_{14}$H$_{10}$,
in its optimized geometry. The choice of C$_{14}$H$_{10}$ as the
monomer is based upon the fact that it has the same chemical formula
as of the repeat unit shown in Fig. \ref{fig:repeat_units}. Moreover,
in experimental situation also PDPA derivatives are usually synthesized
using diphenyl acetylene monomer and the TaCl5-Bu4Sn catalyst.\citet{gontia_prb66_2002}
Therefore, we believe that our choice of using diphenyl acetylene
as a monomer is on a strong footing from a chemical point of view.

Similarly, we define an oligomerization energy, $E_{oligo}$, as \begin{equation}
E_{oligo}=\frac{1}{n}\left(E_{oligomer}-n\ E_{monomer}-2\ E_{H}\right)\label{eq:oligomerization_energy}\end{equation}
 where $E_{oligomer}$ is the total energy of the respective oligomer
and $E_{H}$ is that of the hydrogen atom.

\section{Results\label{sec:Results}}

\subsection{Structural Properties}

\label{sub-sec:structure}

\begin{figure}[ht]
 \includegraphics[width=8cm]{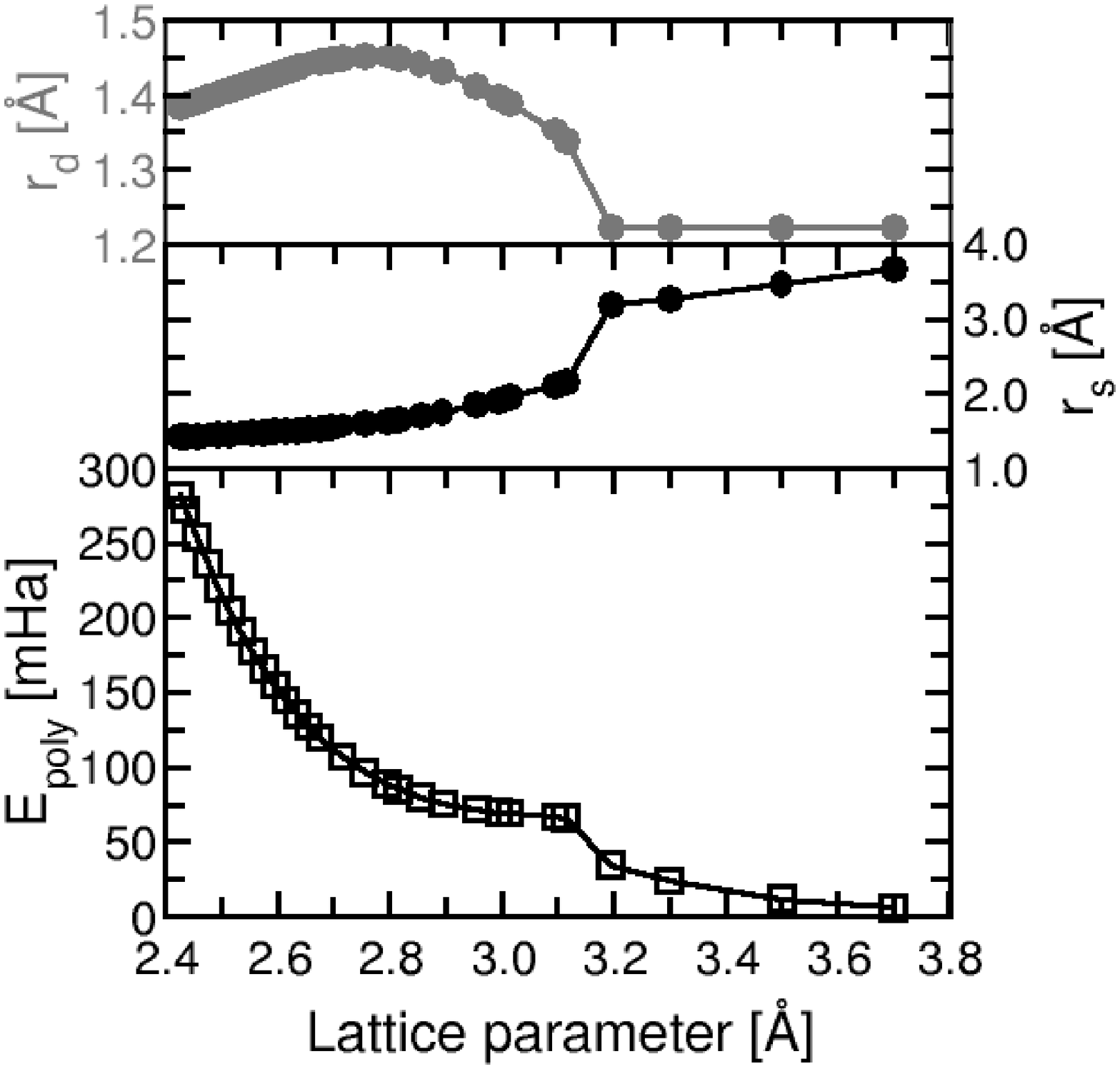}

\caption{Variation of the single bond length, $r_{s}$, double bond length,
$r_{d}$, (top) and polymerization energy, $E_{poly}$, of infinite
PDPA (bottom) as a function of the lattice parameter $a$. }

\label{fig:PDPA_E_poly} 
\end{figure}

To evaluate the stability of PDPA we calculate the polymerization
energy as a function of the lattice parameter $a$, which is displayed
in Fig. \ref{fig:PDPA_E_poly} (bottom panel). The most striking result
is that the values are positive over the whole range, indicating that
the system is not bound at all. The energy monotonically decreases
with increasing lattice parameter up to $a\approx$ 3.1 \AA{}\
followed by a double kink which is explained by a change in the bonding
characteristics. On analyzing the single and double bond lengths (top
panel) we see that first both of them increase due to steric hindrance
between the phenyl rings, until the double bond reaches a value of
$\approx$ 1.45 \AA{}. With even larger lattice spacing, the single
bond length keeps increasing while the double bond length goes down,
because any further increase would correspond to a single bond, which
is impossible given the chemical structure. Once $a$ reaches 3.115
\AA{}, $r_{d}$ becomes smaller than $1.34$ \AA{}, while $r_{s}$
gets larger than $1.54$ \AA{}, values typical for C=C and C-C bonds,
respectively. This finally leads to the breaking of the single bond
and the subsequent transformation of the double bond into a triple
bond, converting the system into 
individual diphenyl acetylene molecules as indicated by the fact that
the binding energy is approaching zero.

\begin{figure}
\includegraphics[width=8cm]{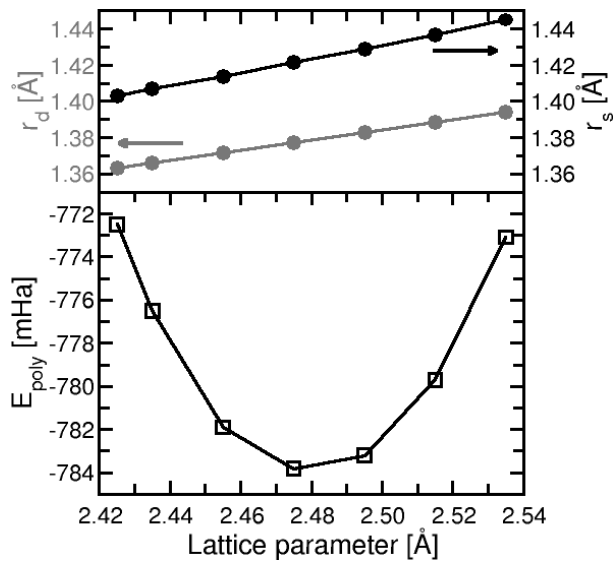}

\caption{Variation of the single bond length, $r_{s}$, double bond length,
$r_{d}$, (top) and polymerization energy, $E_{poly}$, of polyacetylene,
as a function of the lattice parameter $a$. }

\label{fig:PA_E_poly} 
\end{figure}

For comparison, we perform the same calculation for PA. Kahlert and
co-workers\citet{kahlert1987} measured the lattice parameter as well
as C=C and C--C bond lengths to be 2.455 \AA{}, 1.36 \AA{}, and 1.45
\AA{}, respectively, while Yannono et al.\citet{yannoni1983} reported
on the basis of NMR results double and single bond lengths of 1.36
\AA{}\ and 1.44 \AA{}, respectively. Our results, shown in Fig. \ref{fig:PA_E_poly},
demonstrate, as expected, that the system is bound, with the theoretical
lattice parameter being 2.475 \AA{}, and the corresponding values
of $r_{s}$ = 1.42 \AA{}, $r_{d}$ = 1.38 \AA{}. Both, theory and
experiment exhibit nonzero bond alternation, which is fully consistent
with Peierl's theorem,\citet{peierls_theorem} stating that one-dimensional
metals undergo a lattice distortion to minimize their energies, thus
becoming band-gap materials. However, the magnitude of bond alternation,
$\Delta r=0.04$ \AA{}, found by our calculations is roughly a factor
of two smaller than the experimental value.\citet{kahlert1987,yannoni1983}
This underestimation of bond alternation by DFT calculations is a
well-known problem, discussed earlier by various authors.\citet{pickettvogl1989}

Realizing that the PDPA polymer is not stable, the question arises
what limits the length of its oligomers. To this extent we study the
energetics as a function of molecular size, considering two to six
repeat units. For each of these structures, the geometry is optimized
and the results are summarized in Table \ref{tab:structures}. We
note that all the oligomers are found to be invariant under inversion
symmetry, exhibiting point group $C_{i}$.

\begin{figure}[h]
 \vspace{1.5cm}
 \includegraphics[width=8cm]{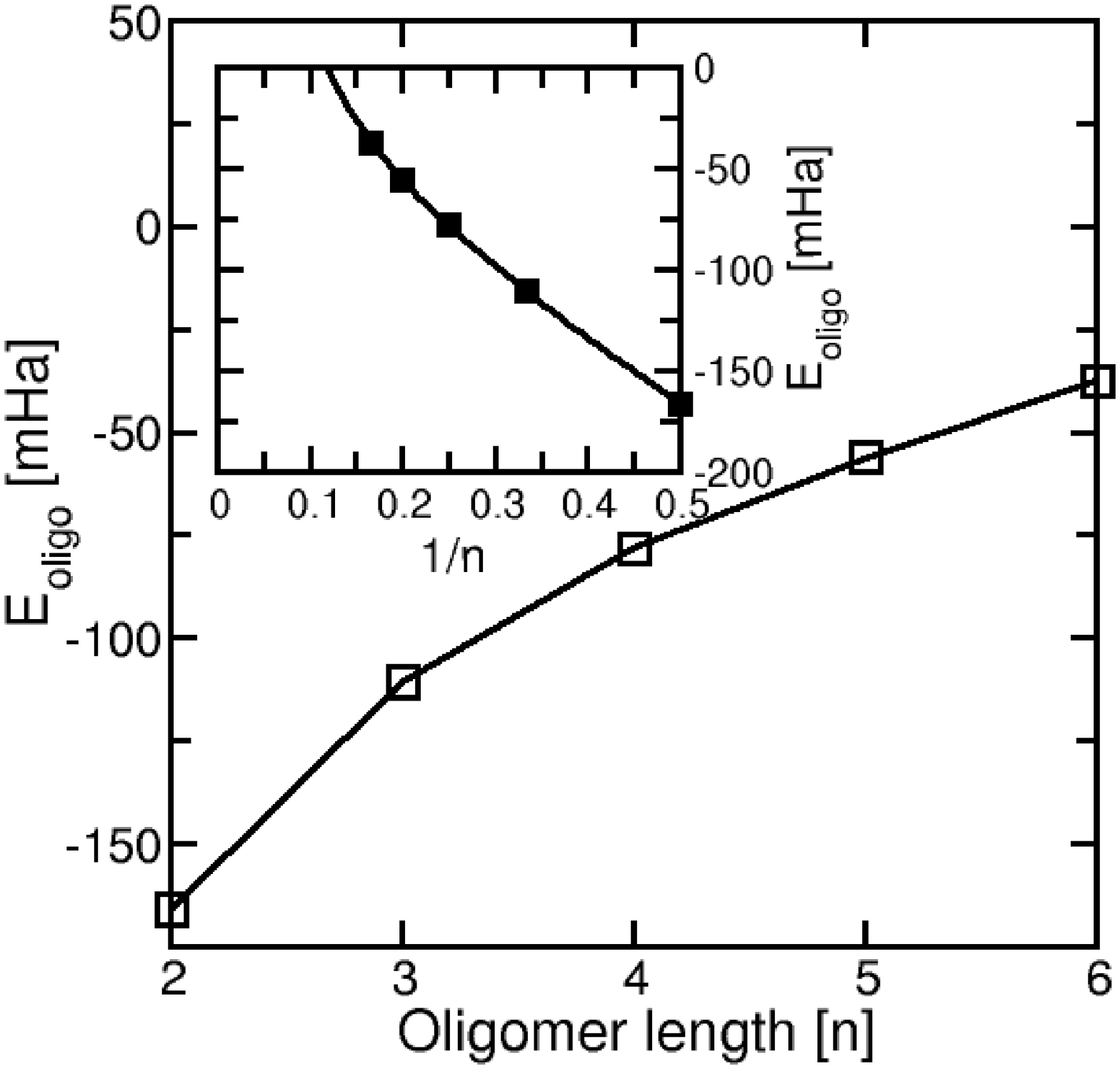}

\caption{Oligomerization energy of PDPA as a function of oligomer length. The
inset presents the same data with respect to the inverse number of
repeat units. The full line represents a cubic-spline fit to the calculated
values, indicating that structures longer than eight repeat units
will be unstable.}

\label{fig:PDPA_E_oligo} 
\end{figure}

Their oligomerization energies, $E_{oligo}$, as defined earlier,
are displayed for these representatives in Fig. \ref{fig:PDPA_E_oligo},
exhibiting a decrease with increasing molecular size. To highlight
the instability with larger oligomer length, we have plotted the energy
as a function of inverse length, i.e., $1/n$ in the inset of the
figure. Here, extrapolation clearly indicates that the formation energy
will become positive for $n>8$. This is in excellent agreement with
experimental findings,\citet{Fujii2001} where the longest oligomers
have been estimated to consist of not more than seven repeat units.

It will be interesting to perform these calculations also for PDPA-7,
however, given the computational constraints, at present it is not
feasible.

\begin{table*}[ht]
 \caption{Structural parameters for the oligomers of PDPA and PA, respectively.
For definitions of the angles see Figure \ref{fig:repeat_units}.
Bond lengths $r_{d}$ and $r_{s}$ as well as bond alternations $\Delta r$
are given in \AA{}. The superscripts denote the index of the repeat
unit.}

\begin{tabular}{|c|c|c|c|c|c||c|c|c|c|c|}
\hline 
 & PDPA-2  & PDPA-3  & PDPA-4  & PDPA-5  & PDPA-6  & PA-2  & PA-3  & PA-4  & PA-5  & PA-6\tabularnewline
\hline 
$r_{d}^{1}$  & 1.37  & 1.38  & 1.38  & 1.38  & 1.39  & 1.35  & 1.35  & 1.35  & 1.35  & 1.35 \tabularnewline
$r_{d}^{2}$  & 1.37  & 1.40  & 1.41  & 1.42  & 1.43  & 1.35  & 1.36  & 1.37  & 1.37  & 1.37 \tabularnewline
$r_{d}^{3}$  &  & 1.38  & 1.41  & 1.43  & 1.44  &  & 1.35  & 1.37  & 1.37  & 1.37 \tabularnewline
$r_{d}^{4}$  &  &  & 1.38  & 1.42  & 1.44  &  &  & 1.35  & 1.37  & 1.37 \tabularnewline
$r_{d}^{5}$  &  &  &  & 1.38  & 1.43  &  &  &  & 1.35  & 1.37 \tabularnewline
$r_{d}^{6}$  &  &  &  &  & 1.39  &  &  &  &  & 1.35\tabularnewline
\hline 
$r_{s}^{1}$  & 1.48  & 1.49  & 1.50  & 1.51  & 1.52  & 1.45  & 1.44  & 1.44  & 1.44  & 1.44 \tabularnewline
$r_{s}^{2}$  &  & 1.49  & 1.51  & 1.53  & 1.55  &  & 1.44  & 1.44  & 1.43  & 1.43 \tabularnewline
$r_{s}^{3}$  &  &  & 1.50  & 1.53  & 1.56  &  &  & 1.44  & 1.43  & 1.43 \tabularnewline
$r_{s}^{4}$  &  &  &  & 1.51  & 1.55  &  &  &  & 1.44  & 1.43 \tabularnewline
$r_{s}^{5}$  &  &  &  &  & 1.52  &  &  &  &  & 1.44\tabularnewline
\hline 
$\Delta r^{1}$  & 0.11  & 0.11  & 0.12  & 0.13  & 0.13  & 0.10  & 0.09  & 0.09  & 0.09  & 0.09 \tabularnewline
$\Delta r^{2}$  &  & 0.09  & 0.10  & 0.11  & 0.12  &  & 0.08  & 0.07  & 0.06  & 0.06 \tabularnewline
$\Delta r^{3}$  &  &  & 0.09  & 0.10  & 0.12  &  &  & 0.07  & 0.06  & 0.06\tabularnewline
$\Delta r^{4}$  &  &  &  & 0.09  & 0.11  &  &  &  & 0.07  & 0.06\tabularnewline
$\Delta r^{5}$  &  &  &  &  & 0.09  &  &  &  &  & 0.07 \tabularnewline
\hline 
$\alpha^{1}$  & 134.0$^{\circ}$  & 138.8$^{\circ}$  & 141.6$^{\circ}$  & 143.3$^{\circ}$  & 143.4$^{\circ}$  & 121.2$^{\circ}$  & 121.1$^{\circ}$  & 121.2$^{\circ}$  & 121.1$^{\circ}$  & 121.3$^{\circ}$\tabularnewline
$\alpha^{2}$  & 113.2$^{\circ}$  & 120.8$^{\circ}$  & 124.7$^{\circ}$  & 126.7$^{\circ}$  & 127.6$^{\circ}$  & 119.5$^{\circ}$  & 118.7$^{\circ}$  & 118.8$^{\circ}$  & 118.8$^{\circ}$  & 118.7$^{\circ}$\tabularnewline
$\alpha^{3}$  &  & 108.3$^{\circ}$  & 115.4$^{\circ}$  & 119.0$^{\circ}$  & 120.1$^{\circ}$  &  & 119.1$^{\circ}$  & 118.5$^{\circ}$  & 118.5$^{\circ}$  & 118.4$^{\circ}$\tabularnewline
$\alpha^{4}$  &  &  & 105.6$^{\circ}$  & 112.6$^{\circ}$  & 115.3$^{\circ}$  &  &  & 119.0$^{\circ}$  & 118.4$^{\circ}$  & 118.4$^{\circ}$\tabularnewline
$\alpha^{5}$  &  &  &  & 104.0$^{\circ}$  & 111.0$^{\circ}$  &  &  &  & 119.0 $^{\circ}$  & 118.3$^{\circ}$\tabularnewline
$\alpha^{6}$  &  &  &  &  & 103.6$^{\circ}$  &  &  &  &  & 119.0$^{\circ}$\tabularnewline
\hline 
$\beta^{1}$  & 112.2$^{\circ}$  & 109.2$^{\circ}$  & 107.2$^{\circ}$  & 105.9$^{\circ}$  & 105.9$^{\circ}$  & 117.1$^{\circ}$  & 117.2$^{\circ}$  & 117.1$^{\circ}$  & 117.1$^{\circ}$  & 117.1$^{\circ}$\tabularnewline
$\beta^{2}$  & 119.8$^{\circ}$  & 112.7$^{\circ}$  & 108.9$^{\circ}$  & 106.5$^{\circ}$  & 105.7$^{\circ}$  & 116.4$^{\circ}$  & 116.8$^{\circ}$  & 117.0$^{\circ}$  & 116.9$^{\circ}$  & 117.1$^{\circ}$\tabularnewline
$\beta^{3}$  &  & 125.3$^{\circ}$  & 117.2$^{\circ}$  & 112.9$^{\circ}$  & 110.8$^{\circ}$  &  & 116.3$^{\circ}$  & 117.0$^{\circ}$  & 117.2$^{\circ}$  & 117.3$^{\circ}$ \tabularnewline
$\beta^{4}$  &  &  & 128.20$^{\circ}$  & 120.1$^{\circ}$  & 115.8$^{\circ}$  &  &  & 116.4$^{\circ}$  & 117.0$^{\circ}$  & 117.1$^{\circ}$\tabularnewline
$\beta^{5}$  &  &  &  & 129.8$^{\circ}$  & 121.6$^{\circ}$  &  &  &  & 116.5$^{\circ}$  & 117.2$^{\circ}$\tabularnewline
$\beta^{6}$  &  &  &  &  & 130.3$^{\circ}$  &  &  &  &  & 116.4$^{\circ}$\tabularnewline
\hline 
$\gamma^{1}$  & 113.8$^{\circ}$  & 112.0$^{\circ}$  & 111.2$^{\circ}$  & 110.8$^{\circ}$  & 110.7$^{\circ}$  & 121.7$^{\circ}$  & 121.7$^{\circ}$  & 121.7$^{\circ}$  & 121.7$^{\circ}$  & 121.6$^{\circ}$ \tabularnewline
$\gamma^{2}$  & 127.0$^{\circ}$  & 126.5$^{\circ}$  & 126.4$^{\circ}$  & 126.8$^{\circ}$  & 126.7$^{\circ}$  & 124.1$^{\circ}$  & 124.5$^{\circ}$  & 124.2$^{\circ}$  & 124.3$^{\circ}$  & 124.2$^{\circ}$ \tabularnewline
$\gamma^{3}$  &  & 126.4$^{\circ}$  & 127.4$^{\circ}$  & 128.1$^{\circ}$  & 129.1$^{\circ}$  &  & 124.6$^{\circ}$  & 124.5$^{\circ}$  & 124.3$^{\circ}$  & 124.3$^{\circ}$\tabularnewline
$\gamma^{4}$  &  &  & 136.4$^{\circ}$  & 127.3$^{\circ}$  & 128.9$^{\circ}$  &  &  & 124.6$^{\circ}$  & 124.6$^{\circ}$  & 124.5$^{\circ}$\tabularnewline
$\gamma^{5}$  &  &  &  & 126.2$^{\circ}$  & 127.4$^{\circ}$  &  &  &  & 124.5$^{\circ}$  & 124.5$^{\circ}$\tabularnewline
$\gamma^{6}$  &  &  &  &  & 126.1$^{\circ}$  &  &  &  &  & 124.6$^{\circ}$\tabularnewline
\hline 
$\delta^{1}$  & 56.3$^{\circ}$  & 65.7$^{\circ}$  & 72.4$^{\circ}$  & 80.6$^{\circ}$  & 78.0$^{\circ}$  &  &  &  &  & \tabularnewline
$\delta^{2}$  & 63.4$^{\circ}$  & 77.1$^{\circ}$  & 85.9$^{\circ}$  & 92.1$^{\circ}$  & 86.9$^{\circ}$  &  &  &  &  & \tabularnewline
$\delta^{3}$  &  & 68.1$^{\circ}$  & 78.0$^{\circ}$  & 84.1$^{\circ}$  & 81.9$^{\circ}$  &  &  &  &  & \tabularnewline
$\delta^{4}$  &  &  & 71.3$^{\circ}$  & 77.8$^{\circ}$  & 79.2$^{\circ}$  &  &  &  &  & \tabularnewline
$\delta^{5}$  &  &  &  & 74.0$^{\circ}$  & 75.8$^{\circ}$  &  &  &  &  & \tabularnewline
$\delta^{6}$  &  &  &  &  & 71.2$^{\circ}$  &  &  &  &  & \tabularnewline
\hline
\end{tabular}\label{tab:structures} 
\end{table*}

Analyzing the structural parameters, we observe a clear trend for
the bond distances. Single as well as double bond lengths not only
increase with the oligomer length, but also for a given oligomer they
increase going from the outer PDPA repeat units toward the innermost
ones. For example, PDPA-6 $r_{d}^{1}$=1.39 \AA{}\ and $r_{d}^{3}$=1.44
\AA{}\ differ by as much as 0.05 \AA{}\, while $r_{s}$ increases
from 1.52 \AA{}\ to 1.56 \AA{}\ at the same time. The latter value
signalizes the bond breaking for larger oligomers as found for the
polymer before. In contrast, for the PA oligomers the corresponding
values all lie in the range of the polymer bond lengths exhibiting
only small variations. The different bond lengths are reflected in
the behavior of the bond alternation along the backbone. For example,
$\Delta r$ for the innermost bonds in PDPA-$6$ is $0.12$ \AA{},
while it is only half as large in PA-6. These results are in clear
contrast to previous model calculations.\citet{shukla_PRL1999}

\begin{figure}[h]
 \includegraphics[width=8cm]{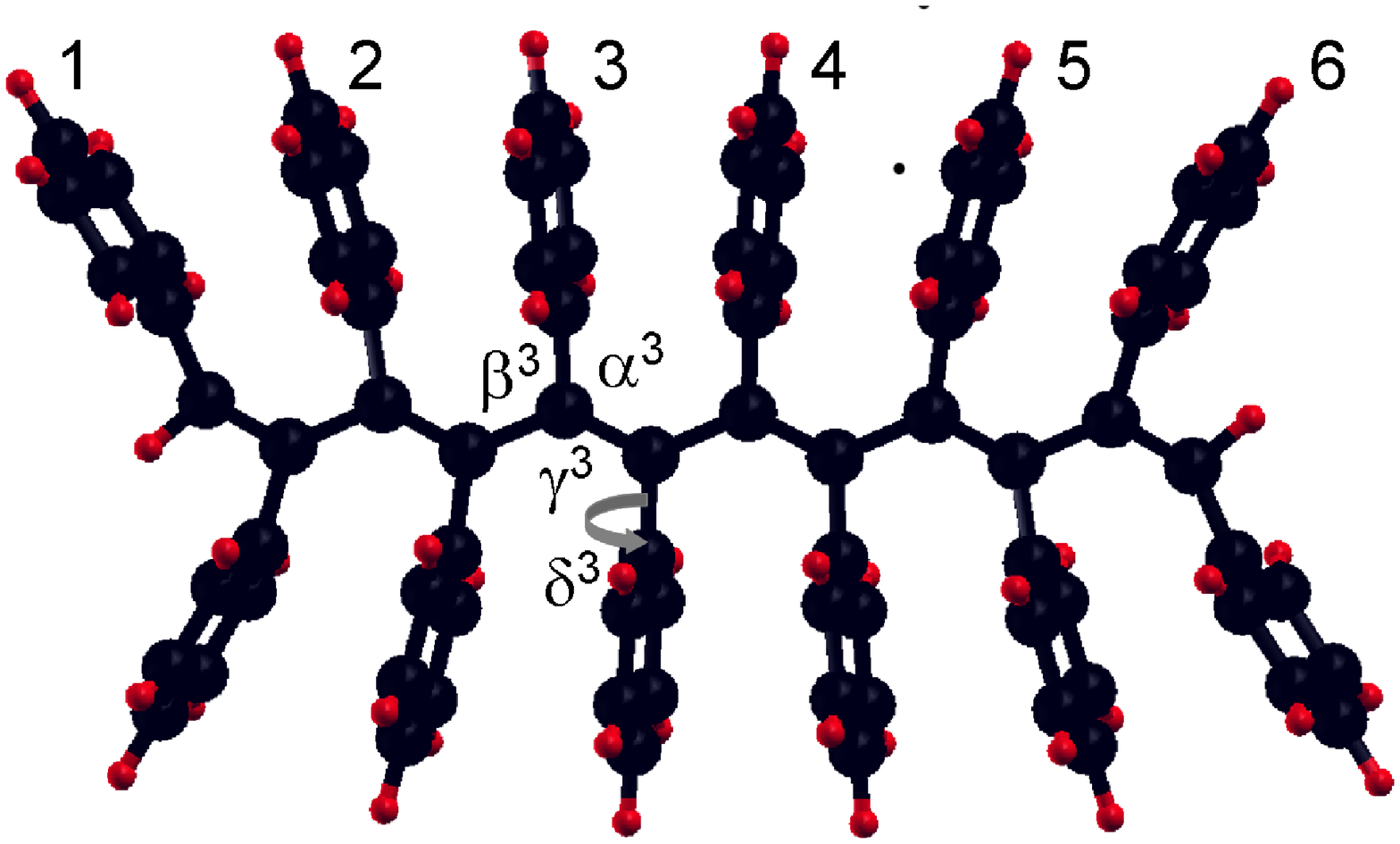}

\caption{(color online) Relaxed geometry of PDPA-6. The angles are exemplary
labeled for the third repeat unit. Note the inversion symmetry of
the system.}

\label{fig:PDPA-6} 
\end{figure}

The instability of longer oligomers is also reflected by the orientation
of the phenyl rings with respect to the backbone, what is evident
from Fig. \ref{fig:PDPA-6} showing the relaxed geometry of PDPA-$6$.
While at the innermost repeat units the phenyl groups take a nearly
standing position with respect to the $x$ axis, the outer ones are
leaning toward the outside. This is demonstrated by the angle $\alpha$
($\beta$), dramatically decreasing (increasing) from the left to
the right.

Another interesting aspect concerns the torsion angle, $\delta$,
of the phenyl ring with respect to the polyene plane. In contrast
to previous work assuming it close to $30^{\circ}$,\citet{shukla_PRL1999}
we predict much larger values, ranging from $56^{\circ}$ to more
than $90^{\circ}$. This appears counter-intuitive at the first sight,
because one would expect that such large torsion angles would destroy
the conjugation between the backbone and the phenyl rings. This fact
is, however, required by the small spacing between the rings provided
by the polyene backbone. Only large torsion angles allow for an accommodation
of the phenyl groups substituting the hydrogen atoms. A similar behavior
is seen in rubrene, in which equally large torsion angles for the
side phenyl rings have been reported.\citet{petrenko2009-rubrene,zhao2006-rubrene}

Before closing the discussion of the structural properties of PDPA's,
we would like to briefly comment on the choice of the exchange-correlation
functional used in these calculations, vis-a-vis some of the hybrid
functionals which have become popular these days.\citet{hybrid-functionals}
These hybrid functionals contain a certain percentage (20\%--40\%)
of exact HF exchange and give quite good results for conjugated polymers
in several cases,\citet{hybrid-functionals} but there is no known
value of the exchange percentage which will give correct results for
all the properties. To investiage as to whehter by using a hybrid
functional one would obtain substantially different results than our
GGA-PBE approach, we performed these structural calculations on PDPA-2
using the PBE0 hybrid functional\citet{PBE0} as implemented in the
ESPRESSO program\citet{espresso} with 25\% HF exchange. The calculation
time for the PDPA-2 molecule with the PBE0 functional increases 400
times, as compared to the GGA-PBE method, thereby making such calculations
impossible for longer oligomers. The structural parameters and the
oligomerization energy of PDPA-2 obtained using these two functionals
are presented in Table \ref{tab:func-compare}, and from the results
it is obvious that most of them agree almost exactly with each other.
The only notable differences are: (a) 3.5\% disagreement in the value
of $\delta^{1}$, and (b) about 9\% difference in $E_{oligo}$, which,
for the present calculations will not alter our conclusions in any
significant manner. We believe that similar trends will also hold
for longer oligomers, and, therefore, the choice of the GGA-PBE approach
in our case is fairly adequate. 
\begin{table}
\caption{Structural parameters and oligomerization energy of PDPA-2 calculated
by GGA-PBE and the PBE0 (hybrid functional) approaches. For definitions
of the angles see Figure \ref{fig:repeat_units}. All the lengths
are in \AA{}, and the energies in milliHartrees.}

\begin{tabular}{|c|c|c|}
\hline 
 & GGA-PBE  & PBE0 \tabularnewline
\hline 
$r_{d}^{1}$  & 1.37  & 1.37 \tabularnewline
$r_{d}^{2}$  & 1.37  & 1.37 \tabularnewline
\hline 
$r_{s}^{1}$  & 1.48  & 1.49 \tabularnewline
\hline 
$\Delta r^{1}$  & 0.11  & 0.12 \tabularnewline
\hline 
$\alpha^{1}$  & 134.0$^{\circ}$  & 133.7$^{\circ}$ \tabularnewline
$\alpha^{2}$  & 113.2$^{\circ}$  & 113.2$^{\circ}$ \tabularnewline
\hline 
$\beta^{1}$  & 112.2$^{\circ}$  & 112.2$^{\circ}$ \tabularnewline
$\beta^{2}$  & 119.8$^{\circ}$  & 119.7$^{\circ}$ \tabularnewline
\hline 
$\gamma^{1}$  & 113.8$^{\circ}$  & 113.7$^{\circ}$ \tabularnewline
$\gamma^{2}$  & 127.0$^{\circ}$  & 127.0$^{\circ}$ \tabularnewline
\hline 
$\delta^{1}$  & 56.3$^{\circ}$  & 58.3$^{\circ}$ \tabularnewline
$\delta^{2}$  & 63.4$^{\circ}$  & 63.7$^{\circ}$ \tabularnewline
\hline 
$E_{oligo}$  & 166.05  & 181.60\tabularnewline
\hline
\end{tabular}\label{tab:func-compare} 
\end{table}

\subsection{Optical Absorption}

\label{sub-sec:optics}

\begin{table}[h]
 \caption{Kohn-Sham band gaps for various oligomers of PDPA and PA.}

\begin{centering}
\begin{tabular}{|c|c|c|c|c|c|}
\hline 
\multicolumn{1}{|c}{} & \multicolumn{5}{c|}{Kohn-Sham gap {[}eV]}\tabularnewline
\hline 
System  & \multicolumn{5}{c|}{$n$}\tabularnewline
\hline 
 & 2  & 3  & 4  & 5  & 6\tabularnewline
\hline 
PA  & 3.88  & 2.96  & 2.40  & 2.02  & 1.74\tabularnewline
\hline 
PDPA  & 2.88  & 2.38  & 2.00  & 1.80  & 1.52\tabularnewline
\hline
\end{tabular}
\par\end{centering}

\label{tab:HOMO--LUMO-gap} 
\end{table}

Having discussed the structural properties, we now have a look at
their impact on the electronic structure. To this extent, we compare
the gap between the highest occupied molecular orbital (HOMO) and
the lowest unoccupied molecular orbital (LUMO) of oligo-PDPA's and
polyenes, presented in Table \ref{tab:HOMO--LUMO-gap}. As expected,
in both the systems, the energy gap decreases with the increase in
molecular length. However, the gap is always found to be smaller in
case of PDPA-$n$ compared to PA-$n$, which is in agreement with
experimental results\citet{tada-fujii-book1997,gontia1999} and earlier
calculations.\citet{shukla_PRL1999} Note that the presented values
are Kohn-Sham gaps with the PBE exchange-correlation functional,\citet{PBE}
which are well known to be underestimated by roughly a factor of two.

\begin{figure}
\includegraphics[width=8cm]{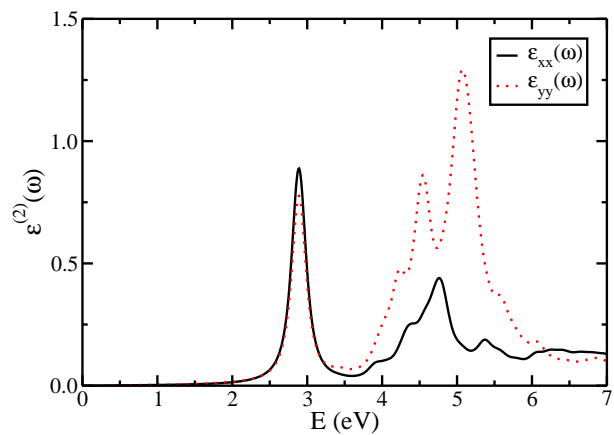}

\caption{(color online) Longitudinal ($\epsilon_{xx}^{(2)}(\omega)$) and transverse components
($\epsilon_{yy}^{(2)}(\omega)$) of the imaginary part of the dielectric
constants of PDPA-2 plotted as a function of the energy of incident
radiation. The calculations were performed using the PBE\citet{PBE}
exchange-correlation functional.\label{Fig-pdpa2-spec}}

\end{figure}

In order to ensure that the Kohn-Sham gaps actually correspond to
the optical gaps, we calculated both the longitudinal and the tranverse
components of the imaginary part of dielectric constant tensor, $\epsilon_{xx}^{(2)}(\omega)$
and $\epsilon_{yy}^{(2)}(\omega)$, respectively ($x$ being the conjugation
direction) for PDPA-2, and the results are presented in Fig. \ref{Fig-pdpa2-spec}.
From the results it is obvious that: (a) the first peaks in both $\epsilon_{xx}^{(2)}(\omega)$
and $\epsilon_{yy}^{(2)}(\omega)$ occur at 2.88 eV, precisely the
Kohn-Sham gap of PDPA-2 reported in Table \ref{tab:HOMO--LUMO-gap},
and (b) significant intensity of the first peak of $\epsilon_{yy}^{(2)}(\omega)$,
indicating a non-trivial transverse polarization component in the
linear absorption at the optical gap, consistent with the earlier
works.\citet{shukla_PRL1999} In order to account for the excitonic
effects in the optical absorption, more sophisticated calculations
including the effects of electron-hole interaction are needed, which
are outside the scope of the present work. Nevertheless, we expect
that in parallel to the gaps, the exciton binding energies will also
decrease with the increasing oligomer length.\citet{Hummer2005,cad2006,Hummer2006}

\begin{centering}
\begin{figure}

\includegraphics[width=6cm]{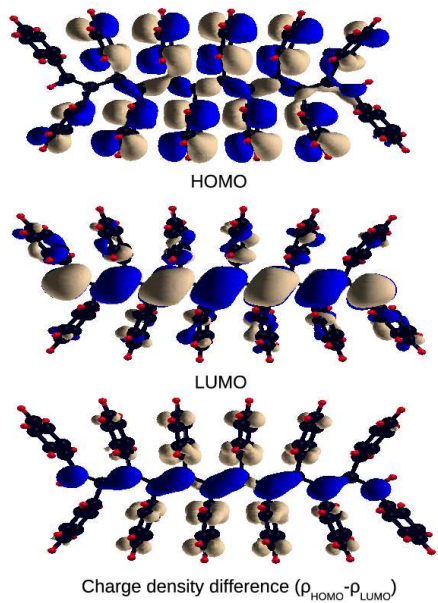}

\caption{(color online) HOMO and LUMO wave functions (isovalue of $\pm0.01$),
and corresponding charge density difference (isovalue of $\pm0.001$
e) plot. Dark (light) regions denote positive (negative) charge. \label{Fig.HOMO_LUMO}}

\end{figure}
\par\end{centering}

Next we examine the the presence of the significant transverse polarization
in the HOMO--LUMO absorption\citet{shukla_PRL1999,Fujii2001} discussed
above, from the point of view of the charge densities of these frontier
orbitals. In light of the large torsion angles of the phenyl rings
this observation needs to be better understood. For this purpose we
present the wave functions of the HOMO and LUMO in Fig. \ref{Fig.HOMO_LUMO}.
While the HOMO has significant charge density, both on the backbone
as well as on the phenyl rings, the LUMO charge is much more confined
to the backbone. This can be better understood when taking the difference
between the two corresponding densities. Thus, HOMO$\rightarrow$LUMO
transitions in the optical spectra should involve significant charge
transfer from the side groups to the backbone atoms, giving rise to
transverse polarization. Moreover, the fact that the HOMO has substantial
charge density on the phenyl rings also indicates that the large torsion
angles do not destroy the electron conjugation between the backbone
and the phenyl rings. A similar trend was also seen by Petrenko \emph{et
al}. for the case of rubrene.\citet{petrenko2009-rubrene}

\section{Conclusions}

In summary, we have demonstrated by total-energy calculations based
on density functional theory that PDPA oligomers are only stable up
to eight repeat units, while the corresponding polymer cannot be formed.
This finding is related to the large size of the phenyl rings which
need to be accommodated within the length of a polyene dimer (repeat
unit of the backbone) confirming and explaining experimental observations.
The torsion angles of the phenyl rings range from $55^{\circ}$ to
$95^{\circ}$, ruling out an earlier conjecture about their values
being $30^{\circ}$ only. At the same time we have shown that such
large torsion angles do not destroy the conjugation of electrons from
the backbone with those of the phenyl rings. This fact is supported
by the electron distribution of the HOMO which exhibits substantial
charge density on the phenyl rings.

The energetics and the resulting structural properties of the the
oligo-PDPAs have been compared in detail with the corresponding situation
in the parent-polymer, polyacetylene. It has been revealed that due
to phenyl substitution, the bond alternation increases while the band
gap decreases. 

\begin{acknowledgments}
This work was supported by the Austrian Science Fund, Project No.
S9714. We thank Peter Blöchl for providing help concerning the PAW
code. 
\end{acknowledgments}

\end{document}